\documentclass[11pt,twoside]{article}

\usepackage{asp2006}
\usepackage{fancyhdr,graphicx,caption,rotating,multirow,lscape}
\DeclareGraphicsExtensions{.eps,.eps.gz,.ps,.jpg,.tiff}

\begin{document}

\title{TrES Exoplanets and False Positives:\\
Finding the Needle in the Haystack}  

\author{F.~T.~O'Donovan}
\affil{California Institute of Technology, 1200 East California Boulevard, Pasadena, CA 91125}
\author{D.~Charbonneau}  
\affil{Harvard--Smithsonian Center for Astrophysics, 60 Garden Street, Cambridge, MA 02138}    

\begin{abstract} 
Our incomplete understanding of the formation of gas giants and of their mass--radius relationship has motivated ground--based, wide--field surveys for new transiting extrasolar giant planets. Yet, astrophysical false positives have dominated the yield from these campaigns. Astronomical systems where the light from a faint eclipsing binary and a bright star is blended, producing a transit--like light curve, are particularly difficult to eliminate. As part of the Trans--atlantic Exoplanet Survey, we have encountered numerous false positives and have developed a procedure to reject them. We present  examples of these false positives, including the blended system \mbox{GSC\,03885--00829} which we showed to be a K dwarf binary system superimposed on a late F dwarf star. This transit candidate in particular demonstrates the careful analysis required to identify astrophysical false positives in a transit survey. From amongst these impostors, we have found two transiting planets. We discuss our follow-up observations of TrES--2, the first transiting planet in the Kepler field. 
\end{abstract}

\section*{Unmasking Impostor Planets}

The Trans--atlantic Exoplanet Survey (TrES) uses a network of three 10\,cm telescopes (Sleuth at Palomar Observatory, California [\citealt{ODonovan_Charbonneau_Kotredes:AIP:2004a}], PSST at Lowell Observatory, Arizona [\citealt{Dunham_Mandushev_Taylor:pasp:2004a}], and STARE at Observatorio del Teide, Tenerife, Spain [\citealt{Alonso_Deeg_Brown:an:2004a}]) to look for stars that show evidence of planetary transits across their disks. Although there are now 14 known transiting planets (two of which, WASP--1 and WASP--2, were announced at this workshop; see \citealt{Collier-Cameron_Bouchy_Hebrard:preprint:2006a}), we still cannot entirely reconcile our observations of the planetary masses and radii with theoretical predictions. For example, four of the known transiting planets, including WASP--1, have radii larger than can be explained by current structural models (see, e.g., \citealt{Bakos_Noyes_Kovacs:preprint:2006a}). We are also undecided as to the formation mechanism for these giant planets, whether it be via core accretion \citep{Pollack:araa:1984a}, or gravitational instability \citep{Boss:sc:1997a}. 

The yield of any ground-based, wide-field transit survey like TrES will be dominated by astrophysical false positives, as there are many astronomical systems involving eclipsing binaries that can reproduce the light curve indicative of a transiting planet. The TrES network observes for several months a $5.7\deg \times 5.7\deg$ field of view containing thousands of nearby bright stars ($9.5\leq V \leq 15.5$). We then examine the resulting light curves for periodic transits with depths of about 1\%, the signature of a transiting gas giant around a dwarf star. However, due to the radius degeneracy of gas giants, brown dwarfs and M-dwarfs, the transits of these objects across a dwarf star can produce the same transit depth. Similarly, a transit of a giant star by a dwarf star may result in a similar photometric signal. If the light from a faint eclipsing binary with a large eclipse depth is blended with that from a brighter star, the resulting dilution may reduce the eclipse depth to a few percent. Every type of astrophysical false positives must be rejected before we can be certain that the transits we observe are from a Jupiter-sized planet. Here we present examples of the types of astrophysical false positives that were rejected along the path to finding TrES-1 \citep{Alonso_Brown_Torres:apjl:2004a} and TrES-2 \citep{ODonovan_Charbonneau_Mandushev:apjl:2006a}. 

\begin{figure}[h]
\centering
\includegraphics[angle=90,width=.65\textwidth]{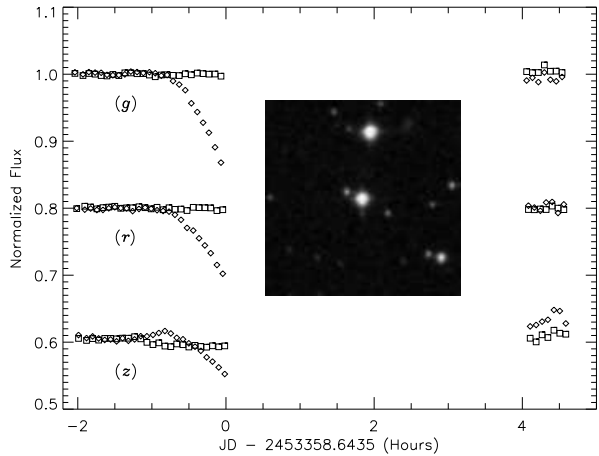}
\caption{(From \citealt{ODonovan_Charbonneau_Alonso:preprint:2006a}.) Follow--up multi-color photometry of a TrES transit candidate (\textit{squares}) and a neighboring star (\textit{diamonds}) within $45\arcsec$. (The inset $2\arcmin \times 2\arcmin$ Digitized Sky Survey image shows both stars with the transit candidate at the center.) Strong winds interrupted the observations before the time of the center of eclipse. However, enough data were obtained to show that the brightness of transit candidate does not vary, whereas the neighboring star displays a deep eclipse. The blending in the TrES observations of the light from this eclipsing binary and the transit candidate resulted in the observed apparent transits. \label{fig:blend}}
\end{figure}

Due to the brightness of our target stars, it is likely that many of them have previously been observed in sky surveys. By comparing these observations with our expectations for a planetary system, we can obtain additional evidence for the nature of each transit candidate. We compare the Tycho--2  $B_{T}-V_{T}$ colors with the visible colors of dwarf stars. Two Micron All Sky Survey (2MASS) infrared ($J-K$) colors can also be used to estimate the stellar radius of our target star (see, e.g., \citealt{Brown:apjl:2003a}). We obtain the proper motions of our transit candidates from the USNO CCD Astrograph Catalog (UCAC2), as nearby dwarf stars should display large proper motions. We also examine the Digitized Sky Survey (DSS) images (see, e.g., Fig.\ref{fig:blend}) of the nearby sky for each transit candidate to identify neighboring stars of similar brightness which might be blended in our lower-angular-resolution TrES images. 

\begin{figure}[h]
\centering
\begin{minipage}[c]{.49\textwidth}
\includegraphics[width=1.\textwidth]{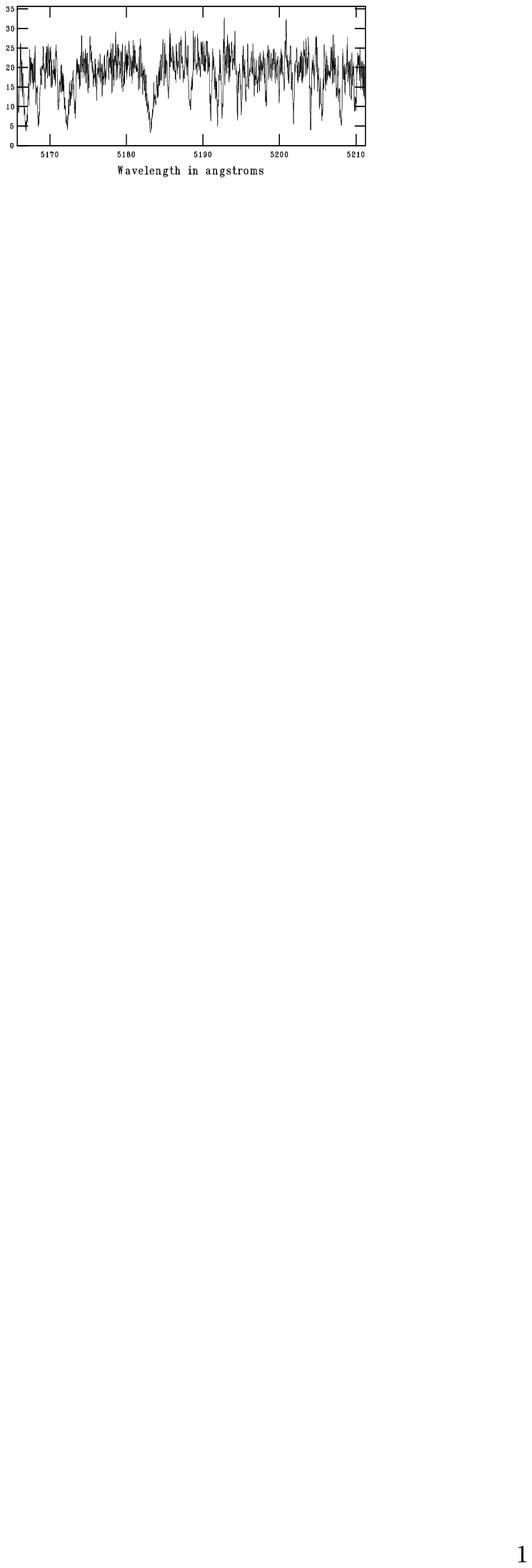}
\end{minipage}
\begin{minipage}[c]{.49\textwidth}
\includegraphics[width=1.\textwidth]{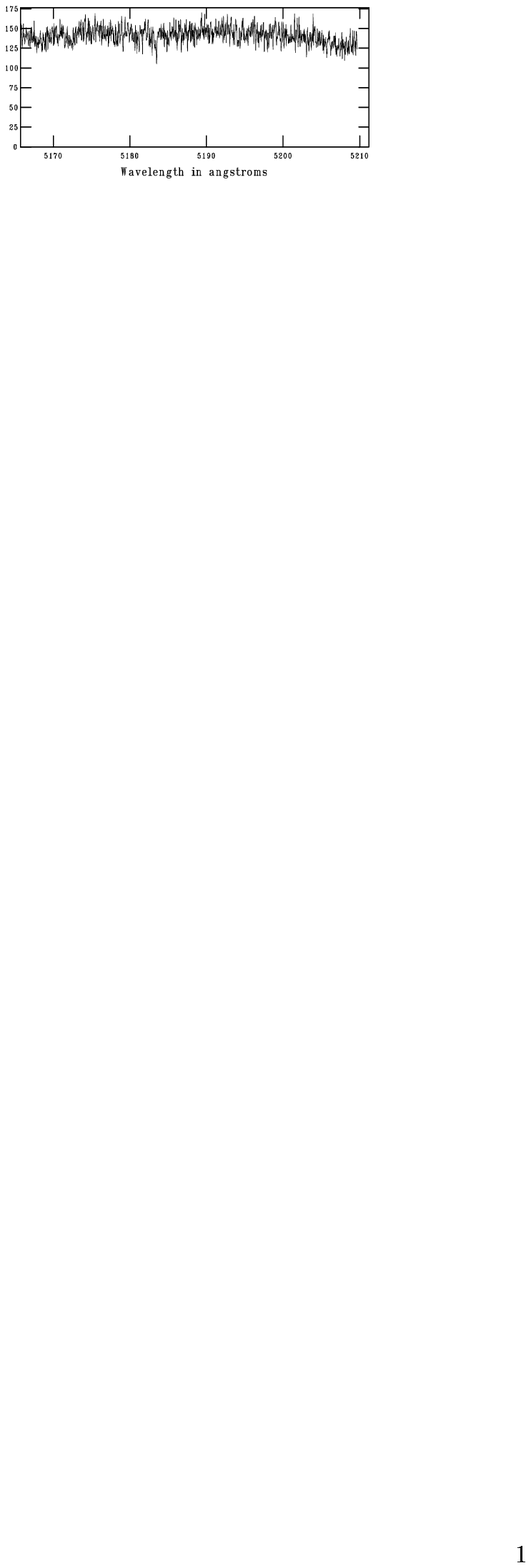}
\end{minipage}
\caption{(From \citealt{ODonovan_Charbonneau_Alonso:preprint:2006a}.) Spectra of two TrES transit candidates observed using the CfA Digital Speedometers. The transit candidate discussed in Figure~\ref{fig:blend} displays the spectrum on the left, that of an early K--dwarf, The spectrum on the right is a featureless spectrum of a A--type star, hence we rejected this candidate. \label{fig:spectra}}
\end{figure}

All of our TrES transit candidates are spectroscopically monitored with the Harvard--Smithsonian Center for Astrophysics Digital Speedometers \citep{Latham:ASP:1992a}. Examples of the spectra obtained are given in Figure~\ref{fig:spectra}. Usually, from a single spectrum such as these, we can determine the spectral type and luminosity class of each transit candidate, and eliminate those with the large stellar radii of stars earlier than F. We then obtain additional observations of the remaining candidates, from which we measure the radial velocity variations of the stars. This allows us to identify velocity variations due to companions of masses greater than $\sim10~\mathrm{M_{Jup}}$ for short-period systems. 

Although we can obtain a radius estimate from our TrES observations of a transit candidate, we must obtain high-precision photometry in order to get a precise radius measurement that is useful for comparison with theoretical predictions. Ideally, this high-precision photometry would be obtained during the same observing season as the TrES observations, while the object can still be observed for $\sim6$ hours each night. In reality, the difficulty in reducing the large TrES data set, identifying candidates and scheduling observing time on large telescopes prohibits this in the majority of the cases. Here, spectroscopic followup has the advantage that spectra of the target can be obtained much later in the season. One consequence of postponing photometric followup is that we rely heavily on the accuracy of the orbital ephemeris we derive from our TrES observations. If we do not have an accurate ephemeris, we may not be able to recover the transits of a transiting planet. In the case of the transit candidate discussed in Figure~\ref{fig:blend}, we obtained follow-up photometry (see Fig.~\ref{fig:blend}) with the FLWO 1.2-m telescope on Mt.~Hopkins, Arizona. Although weather prevented full coverage of the predicted time of transit, we did observe at the time of ingress. We did not see any variation in the flux from the transit candidate, which might suggest an inaccurate prediction. However, a neighboring star was found to be an eclipsing binary with an eclipse at the predicted transit time, and an orbital period equal to that of the transit candidate. The observed transits of the transit candidate were therefore the result of a blend of the target star and the nearby eclipsing binary in our TrES observations.

\begin{figure}[h]
\centering
\includegraphics[width=.50\textwidth]{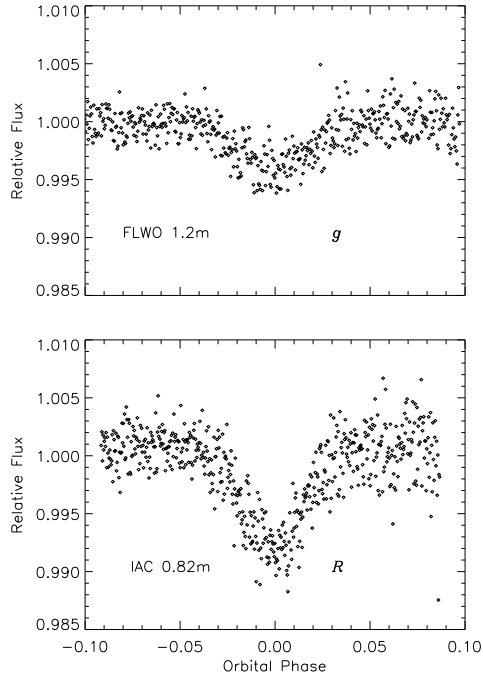}
\caption{(From \citealt{ODonovan_Charbonneau_Torres:apj:2006a}.) Multi-color photometry of the TrES transit candidate \mbox{GSC\,03885--00829} showing the color-dependent eclipse depths indicative of a blended eclipsing binary. \label{fig:eclipses}}
\end{figure}

Follow-up photometry can also reveal blends through the use of multiple filters. When the transits of a candidate are recovered using multi-color photometry, we can compare the transit depths as a function of observed wavelengths. If the transit is the result of a dark planet passing in front of a star, the transit depth should be color-independent. However, in the case of some blends, where the stellar color of the brightest star is different from that of the primary star of the binary, the eclipse depths will vary with wavelength. For example, the TrES candidate \mbox{GSC\,03885--00829} \citep{ODonovan_Charbonneau_Torres:apj:2006a} passed our initial spectroscopic tests. However, our multi-color followup (see Fig.~\ref{fig:eclipses}) obtained color-dependent eclipse depths, indicating the observed transits were in fact the result of a blend. We were able to construct a blend model of a K dwarf binary system and on a late F dwarf star that reproduced our observations, including the original TrES observations, the lack of radial velocity variations on the stellar mass scale and the color-dependent eclipse depths.

\section*{TrES--2: The Most Massive Nearby Transiting Planet}

\begin{figure}[h]
\centering
\includegraphics[width=.55\textwidth]{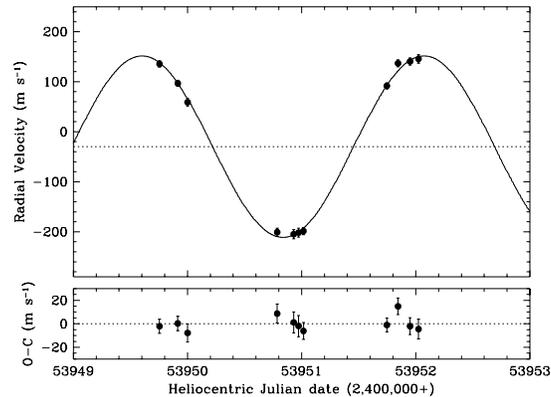}
\caption{(From \citealt{ODonovan_Charbonneau_Mandushev:apjl:2006a}.) Keck/HIRES radial velocity observations of the transiting planet TrES--2. The best-fit orbit and $\gamma$-velocity are overplotted, and the residuals to the best-fit are shown below. These resource-intensive observations are the final step in confirming a transiting planet, hence the majority of astrophysical false positives must be rejected from a transit candidate list before obtaining such observations. \label{fig:rv}}
\end{figure}

\begin{figure}[h]
\centering
\includegraphics[width=.75\textwidth]{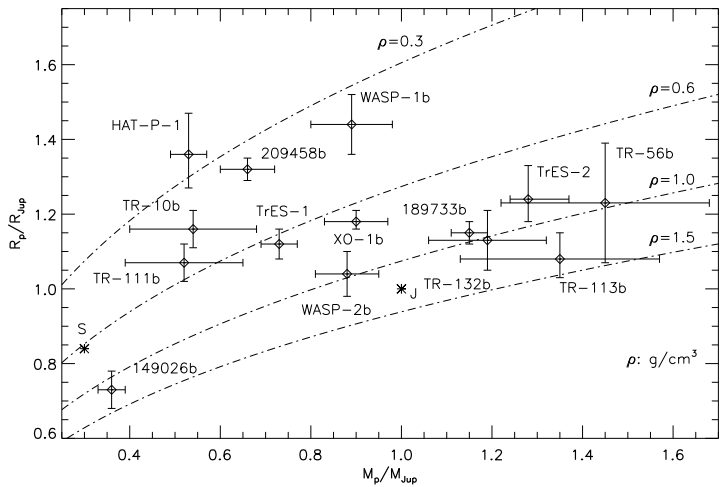}
\caption{Masses and radii for the 14 known transiting extrasolar planets. Jupiter (J) and Saturn (S) are also plotted for comparison. Although \mbox{HD\,209458\,b} was initially believed to be the only known transiting planet with a radius exceeding model predictions, three planets announced recently have also displayed large radii. \label{fig:rm}}
\end{figure}

Although it is possible to cull most of the astrophysical false positives from our list of transit candidates through the use of follow-up spectroscopy and photometry, we cannot be certain of the true nature of the remaining candidates without obtaining the spectroscopic orbit due to the planet. This requires high-resolution spectroscopy with a high signal-to-noise ratio using very large telescopes. An example of such observations is given in Figure~\ref{fig:rv}, which shows the radial velocity variations of the star TrES--2 caused by the presence of the $1.3$-$\mathrm{M_{Jup}}$ planet TrES--2 \citep{ODonovan_Charbonneau_Mandushev:apjl:2006a}. The planet TrES--2 is the most massive transiting planet within 300\,pc, and is one of four planets that have radii larger than can be explained using current structural models (see Fig.~\ref{fig:rm}; \citealt{Laughlin_Wolf_Vanmunster:apj:2005a} and \citealt{Charbonneau_Brown_Burrows:PPV:2006a} give a summary of the attempts to explain the inflated radius of one of these planets, \mbox{HD\,209458\,b}). Since three of these four planets were announced within months of this workshop, the field of transiting exoplanets has had a rapid influx of new data with which to improve our theoretical models of these gas giants. 

Although TrES--2 was announced only in September 2006, it has already been the subject of detailed studies. As reported by M.~Holman in these proceedings, high-precision photometry of multiple transits of TrES--2 was obtained as part of of the Transit Light Curve project. An analysis of the timing of these transits may indicate the presence of additional planets in the TrES--2 system that perturb the orbit of TrES--2. This is of particular importance to prepare for {\it Kepler} observations of this planet, which is the first known transiting planet in the {\it Kepler} field. By the time of publication of these proceedings, {\it Spitzer} IRAC observations of TrES--2 during a secondary eclipse should be complete. From these data, we will determine the planetary flux at infrared wavelengths, and place constraints upon the atmospheric abundances of the molecules (CH$_{4}$, CO  and H$_{2}$O) that dominate the planetary spectrum at those wavelengths (see, e.g., \citealt{Charbonneau_Brown_Burrows:PPV:2006a}).

With the recent significant increase in the number of known transiting planets, it appears that several transit survey teams have developed extensive experience in weeding out astrophysical false positives from their transit candidate lists and in adequately confirming the true nature of each possible planetary system. We look forward to the results of intensive study of the new transiting planets, including TrES--2, which will surely change our view of extrasolar planets before the next Transiting Extrasolar Planets Workshop. 

\acknowledgements 
We thank Lynne Hillenbrand for her continued support of this thesis work. This material is based on work supported by the National Aeronautics and Space Administration under grant NNG05GJ29G, issued through the Origins of Solar Systems Program.


\begin{thebibliography}{}

\bibitem[\protect\citeauthoryear{{Alonso} et~al.}{{Alonso}
  et~al.}{2004a}]{Alonso_Deeg_Brown:an:2004a}
{Alonso}, R., et~al. 2004a, Astron. Nachr., 325, 594

\bibitem[\protect\citeauthoryear{{Alonso} et~al.}{{Alonso}
  et~al.}{2004b}]{Alonso_Brown_Torres:apjl:2004a}
{Alonso}, R., et~al. 2004b, ApJ, 613, L153

\bibitem[\protect\citeauthoryear{{Bakos} et~al.}{{Bakos}
  et~al.}{2006}]{Bakos_Noyes_Kovacs:preprint:2006a}
{Bakos}, G.~A., et~al. 2006, ApJ, in press (astro-ph/0609369)

\bibitem[\protect\citeauthoryear{{Boss}}{{Boss}}{1997}]{Boss:sc:1997a}
{Boss}, A.~P. 1997, Science, 276, 1836

\bibitem[\protect\citeauthoryear{{Brown}}{{Brown}}{2003}]{Brown:apjl:2003a}
{Brown}, T.~M. 2003, ApJ, 593, L125

\bibitem[\protect\citeauthoryear{{Charbonneau} et~al.}{{Charbonneau}
  et~al.}{2006}]{Charbonneau_Brown_Burrows:PPV:2006a}
{Charbonneau}, D., et~al. 2006, in press, (astro-ph/0603376)

\bibitem[\protect\citeauthoryear{{Collier Cameron} et~al.}{{Collier Cameron}
  et~al.}{2006}]{Collier-Cameron_Bouchy_Hebrard:preprint:2006a}
{Collier Cameron}, A., et~al. 2006, MNRAS, submitted (astro-ph/0609688)

\bibitem[\protect\citeauthoryear{{Dunham} et~al.}{{Dunham}
  et~al.}{2004}]{Dunham_Mandushev_Taylor:pasp:2004a}
{Dunham}, E.~W., et~al. 2004, PASP, 116, 1072

\bibitem[\protect\citeauthoryear{{Latham}}{{Latham}}{1992}]{Latham:ASP:1992a}
{Latham}, D.~W. 1992, in ASP Conf. Ser. 32, IAU Colloq. 135, 110

\bibitem[\protect\citeauthoryear{{Laughlin} et~al.}{{Laughlin}
  et~al.}{2005}]{Laughlin_Wolf_Vanmunster:apj:2005a}
{Laughlin}, G., et~al. 2005, ApJ, 621, 1072

\bibitem[\protect\citeauthoryear{{O'Donovan}, {Charbonneau}, \&
  {Kotredes}}{{O'Donovan}
  et~al.}{2004}]{ODonovan_Charbonneau_Kotredes:AIP:2004a}
{O'Donovan}, F.~T., {Charbonneau}, D.,  \& {Kotredes}, L. 2004, in AIP Conf.
  Proc. 713, 169

\bibitem[\protect\citeauthoryear{{O'Donovan} et~al.}{{O'Donovan}
  et~al.}{2006a}]{ODonovan_Charbonneau_Mandushev:apjl:2006a}
{O'Donovan}, F.~T., et~al. 2006a, ApJ, 651, L61

\bibitem[\protect\citeauthoryear{{O'Donovan} et~al.}{{O'Donovan}
  et~al.}{2006b}]{ODonovan_Charbonneau_Alonso:preprint:2006a}
{O'Donovan}, F.~T., et~al. 2006b, ApJ, submitted (astro-ph/0610603)

\bibitem[\protect\citeauthoryear{{O'Donovan} et~al.}{{O'Donovan}
  et~al.}{2006c}]{ODonovan_Charbonneau_Torres:apj:2006a}
{O'Donovan}, F.~T., et~al. 2006c, ApJ, 644, 1237

\bibitem[\protect\citeauthoryear{{Pollack}}{{Pollack}}{1984}]{Pollack:araa:198%
4a}
{Pollack}, J.~B. 1984, ARA\&A, 22, 389

\end{thebibliography}
\end{document}